\documentclass{ws-procs9x6}

\begin{document}

\title{Azimuthally-sensitive Pion HBT at RHIC}

\author{Mike Lisa, for the STAR Collaboration}

\address{
Department of Physics, Ohio State University, Columbus, OH 43210, USA
}

\maketitle

\abstracts{
The STAR Collaboration has measured 
two-pion 
correlation functions versus emission angle with
respect to the event plane in non-central Au+Au collisions at
$\sqrt{s_{NN}}=130,200$~GeV.  In the context of a parameterized freezout scenario,
the data suggest an out-of-plane-extended
freezeout geometry, and a rapid system evolution to freezeout.
}

Studies of nuclear collisions at the Relativistic Heavy Ion Collider (RHIC) challenge
our understanding of strongly-interacting systems under extreme conditions.
Momentum-space observables in the soft ($p_T<2$~GeV) sector
have been successfully reproduced
by hydrodynamical models\cite{KHHH01,STARv2ID,TLS01},
which should 
approximate the system evolution in the
high-density phase of the collision.
However, two-particle correlations,
which probe the space-time structure of the system\cite{WH99},
are not well-reproduced by most models.
That theory generates the correct dynamic {\it signatures} (e.g. collective flow) in
momentum space, while apparently
following the wrong space-time dynamics, has been called the ``RHIC HBT puzzle.''\cite{H02_PANIC02}

To model the collision more realistically than with hydrodynamics alone, groups\cite{TLS01,SBD01} have 
implemented hadronic ``afterburners'' to represent the dilute stage of the collision.
In these calculations\cite{TLS01}, the late hadronic
stage has little effect on dynamic momentum-space observables such as elliptic flow 
at RHIC energies; however, discrepancies with measured HBT systematics {\it worsen}
with the inclusion of the hadronic stage.  This may be due to the increased timescale
of particle emission\cite{TLS01}.

If the system created at RHIC is indeed evolving/emitting on a shorter timescale than
our present understanding allows, then this is important information, and
it is crucial that more insight be gained on the space-time dynamics of the system.
Correlating HBT measurements with the event plane in non-central
collisions allows a much more detailed exploration of the system evolution
and interplay between the anisotropic geometry and flow fields\cite{VC96,W98,LHW00,HHLW02}.
It also provides a valuable (though model-dependent) measure of the evolution time of
the system until freezeout,
by comparing the final-state freezeout anisotropic geometry to that of the initial overlap
region between the colliding nuclei.
Finally the role of the late hadronic stage may be explored, as the transverse source
geometry exhibits a {\it qualitative} change from an out-of-plane shape
as calculated by hydrodynamics\cite{TLS01,HK02}, to an in-plane shape when the late hadronic
stage is also considered\cite{TLS01}.

\begin{figure}[b]
\vspace*{-0.8cm}
\begin{minipage}[t]{55mm}
\centerline{\epsfxsize=60mm\epsfbox{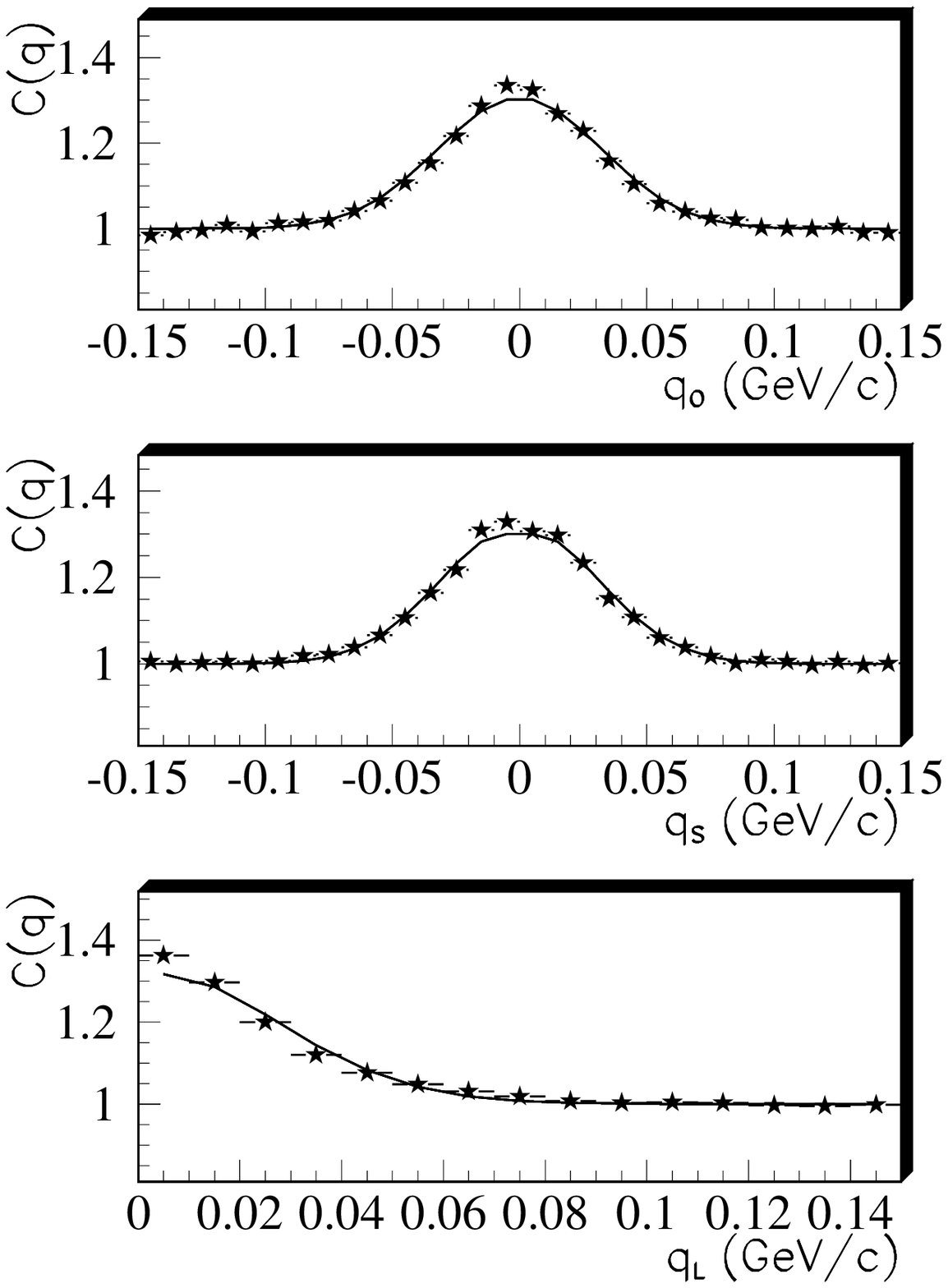}}   
\caption{
Stars show measured one-dimensional projections of the correlation
function for $\phi=45^\circ$, for Au+Au collisions at $\sqrt{s_{NN}}=130$~GeV,
integrated over 30 MeV/c in the unplotted components.
 Lines show projections of the
Gaussian fit.
\label{fig:projections}
}
\end{minipage}
\hspace{\fill}
\begin{minipage}[t]{55mm}
\centerline{\epsfxsize=55mm\epsfbox{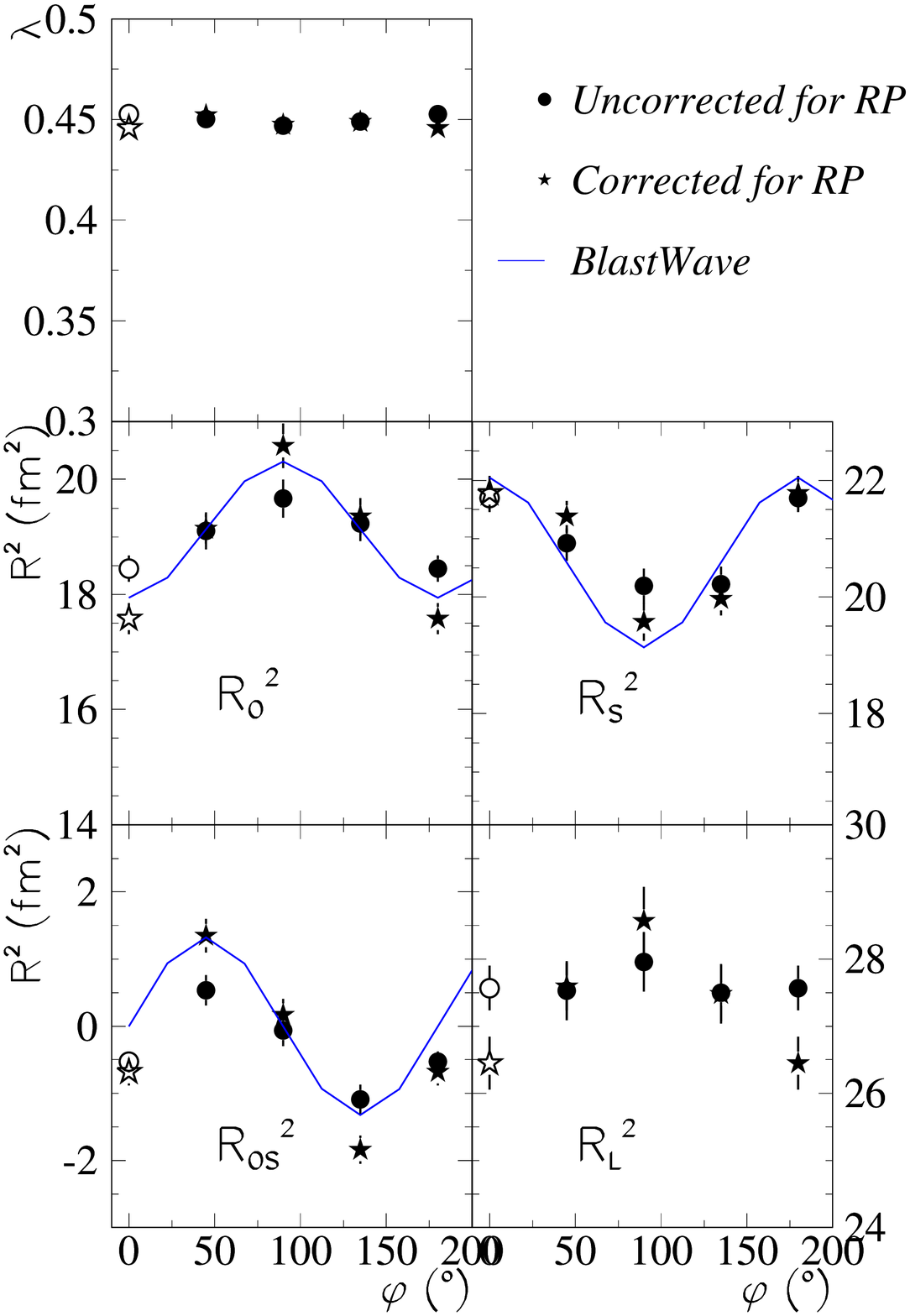}}
\caption{
HBT parameters for Au+Au collisions at $\sqrt{s_{NN}}=130$~GeV.
Stars (circles) show fit parameters to correlation functions corrected (uncorrected)
for finite event-plane resolution and binning effects.  Curves indicate
blast-wave calculations.
\label{fig:Y1radii}}
\end{minipage}
\end{figure}

Results and analysis details from
azimuthally-integrated HBT in STAR have been reported previously\cite{STARHBT}.
The principle differences in the present analysis are: 1) for the construction
of the ``background'' distribution, event-mixed pairs are
formed only from events with similar event-plane orientations in the lab;
2) separate correlation functions are formed for cuts on pair emission
angle with respect to the event plane; 3) prior to fitting the correlation functions,
the ``real'' and ``background'' distributions
are corrected\cite{HHLW02} for effects of finite event-plane resolution and
binning.

The correlation functions, as a function of relative momentum in the Pratt-Bertsch\cite{PrattBertsch}
``out-side-long'' decomposition, were constructed in the Longitudinal Co-Moving System
and fitted with the Gaussian form\cite{WH99}
\begin{equation}
\label{eq:Gaussform}
C(q_o,q_s,q_l,\Phi) = 1 + \lambda(\Phi)
    e^{-q_o^2R_o^2(\Phi)-q_s^2R_s^2(\Phi)-q_l^2R_l^2(\Phi)-2q_oq_lR_{os}^2(\Phi)} ,
\end{equation}
where the HBT parameters depend on $\Phi$, the pair emission angle with respect to the
2$^{\rm nd}$-order event plane\cite{PV98} estimated from the elliptic flow signal.  Since
the 2$^{\rm nd}$-order event plane is used, the HBT radii $R_{ol}^2$ and  $R_{sl}^2$
vanish by symmetry and the remaining radii must show even-order oscillations\cite{HHLW02}.

Correlation functions for Au+Au collisions at $\sqrt{s_{NN}}=130$~GeV were constructed
from  $\sim10^5$ minimum-bias events which passed quality cuts\cite{RandyPhD}.
For statistical reasons, $\pi^-$ and $\pi^+$ data
were combined.
One-dimensional projections of the correlation function for $\Phi=45^\circ\pm22.5^\circ$
are shown in Figure~\ref{fig:projections}, together with fits using Eq.~\ref{eq:Gaussform}.
Fig.~\ref{fig:Y1radii} shows the fit parameters as a function of $\Phi$, both with and without
the correction for event-plane resolution and finite $\Phi$-binning.  The correction
increases the amplitude of the oscillations $R^2_o$, $R^2_s$, and $R^2_{os}$ by $\sim2\times$,
and $R^2_l$ significantly more; the average values of the radii are negligibly affected.

A simple ``blast-wave'' parameterization\cite{FabriceBW} of the freezeout distribution has proven
quite successful at
reproducing $p_T$ spectra\cite{NuKanetaQM01}, the $p_T$-dependence of HBT radii\cite{MercedesQM02},
elliptic flow\cite{STARv2ID}, and correlations between non-identical
particles\cite{FabriceBW,FabriceQM02}.  In particular, the fit to elliptical flow at
$\sqrt{s_{NN}}=130 GeV$\cite{STARv2ID}
suggested a spatial anisotropy for the average source created in minimum-bias collisions.
Indeed, blast-wave fits to the azimuthally-sensitive HBT data {\it require} an out-of-plane
spatial deformation of the source, in agreement with the elliptic flow fits.
It is impossible
in this parameterization to simultaneously reproduce the measured elliptical flow and
azimuthally-sensitive HBT measurements, without an out-of-plane-extended source.  Superimposed
on the transverse HBT radii in Fig.~\ref{fig:Y1radii} are blast-wave calculations
with a 5\% larger source radius out of the event plane than in-plane, and a short ($\sim2$ fm/c)
emission duration.  In addition, the out-of-plane extension of the freezeout distribution points
to a short evolution timescale ($\sim 8-11$~fm/c),
in agreement with simple fits\cite{MS88} to of $R_l(m_T)$.\cite{H02_PANIC02,MercedesQM02}.

\begin{figure}[t]
\centerline{\epsfxsize=110mm\epsfbox{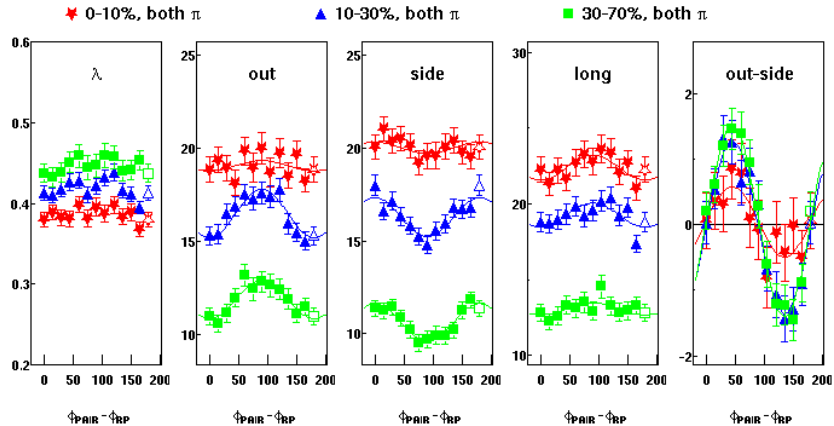}}
\caption{
Preliminary HBT fit parameters
 for Au+Au collisions at $\sqrt{s_{NN}}=200$~GeV.
Stars, triangles, and squares indicate the results for central, midcentral, and peripheral
collisions, respectively.  Lines are fits
to the allowed
2$^{\rm nd}$-order oscillations of the radii.  No correction for event-plane resolution has been applied.
\label{fig:Y2radii}
\vspace*{-3mm}
}
\end{figure}

STAR's higher statistics dataset for $\sqrt{s_{NN}}=200$ GeV collisions will allow more detailed
exploration of the centrality and $k_T$ systematics of azimuthally-sensitive HBT, which encode
important information on the interplay between anisotropic geometry and flow in the collision
dynamics\cite{HK02}.  In Figure~\ref{fig:Y2radii} are plotted preliminary HBT parameters vs. $\Phi$,
for three event centralities.  With 12 $\Phi$-bins, the oscillations are especially convincing, and
we observe the expected result of reduced oscillations for central collisions.

A combined analysis of spectra, elliptic flow, and HBT data is underway, to fully
characterize the freezeout distribution at RHIC.

\end{document}